\documentclass[a4paper,twoside]{article}

\usepackage{epsfig}
\usepackage{subcaption}
\usepackage{calc}
\usepackage{amsmath,amssymb,amsfonts}
\usepackage{amstext}
\usepackage{amsthm}
\usepackage{multicol}
\usepackage{pslatex}
\usepackage{apalike}
\usepackage{multirow}
\usepackage{booktabs}
\usepackage{SCITEPRESS}     

\begin{document}

\title{Analysis of Feature Representations\\ for Anomalous Sound Detection}

\author{\authorname{Robert Müller\orcidAuthor{0000-0003-3108-713X}, Steffen Illium\orcidAuthor{0000-0003-0021-436X}, Fabian Ritz\orcidAuthor{0000-0001-7707-1358} and Kyrill Schmid\orcidAuthor{0000-0001-6748-4896}}
\affiliation{Mobile and Distributed Systems Group, LMU Munich, Germany}
\email{\{robert.mueller, steffen.illium, fabian.ritz, kyrill.schmid\}@ifi.lmu.de}
}


\abstract{In this work, we thoroughly evaluate the efficacy of pretrained neural networks as feature extractors for anomalous sound detection. In doing so, we leverage the knowledge that is contained in these neural networks to extract semantically rich features (representations) that serve as input to a Gaussian Mixture Model which is used as a density estimator to model normality. We compare feature extractors that were trained on data from various domains, namely: images, environmental sounds and music.
Our approach is evaluated on recordings from factory machinery such as valves, pumps, sliders and fans.
All of the evaluated representations outperform the autoencoder baseline with music based representations yielding the best performance in most cases. These results challenge the common assumption that closely matching the domain of the feature extractor and the downstream task results in better downstream task performance.}
\onecolumn \maketitle \normalsize \setcounter{footnote}{0} \vfill
\section{\uppercase{Introduction}}
\label{sec:introduction}
\noindent In the emerging field of anomalous sound detection (ASD), one aims to develop computational methods to reliably detect anomalies in acoustic sounds. These methods can be considered as the counterpart to anomaly detection on visual data and are used in situations where visual monitoring is infeasible. One of the most important use-cases is the early detection of malfunctions during the operation of factory machinery. 
A robust ASD system reduces repair costs, improves safety and prevents consequential damages by enabling early maintenance. Moreover, it reduces the financial burden, an aspect that becomes increasingly important considering the rising costs of modern machinery and equipment.\\ 
\begin{figure}[ht]
  \centering
  \includegraphics[width=0.9\linewidth]{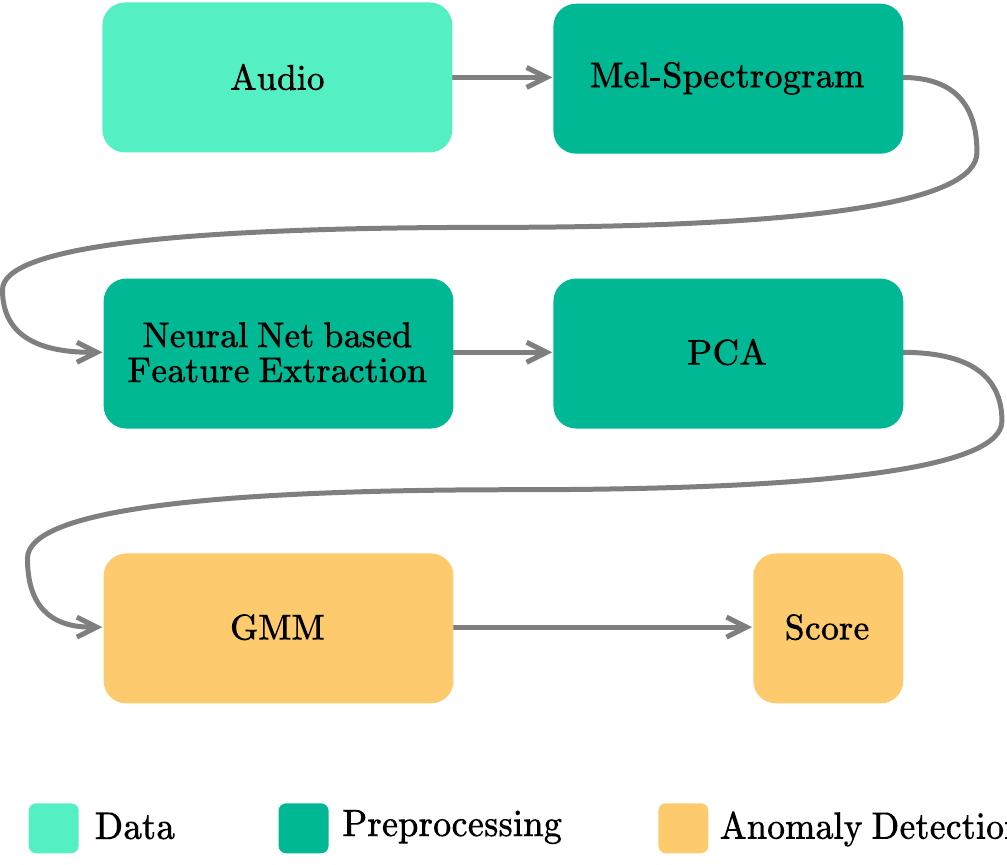}
  \caption{High-level overview of the proposed workflow.}
    \label{fig:workflow}
\end{figure}
While our work focuses on the scenario above, other ASD systems have been developed for closely related applications such as monitoring production processes for irregularities~\cite{hasan2018deep}, the detection of leaks in water supply networks~\cite{Muller2020b} and sound-based security systems for public spaces~\cite{Hayashi2018}.\\
Since it is expensive and tedious to collect an exhaustive number of anomalous samples that cover all possible anomalies in such settings, training ASD models is usually carried out on normal operation data only to learn a notion of normality.
Moreover, even if it was possible to collect large amounts of anomalous data, some anomalies might not even be known beforehand and are thus not available during training.\\
Most of the recent ASD approaches rely on deep autoencoders (AEs). AEs first compress their input into a low dimensional latent code using an encoder neural network (NN). This code is subsequently used to reconstruct the input with a decoder network. Hence, AEs do not use external labels. It is assumed that patterns seen during training yield a lower reconstruction error than those not seen during training. Consequently the reconstruction error is used as a measure of abnormality. However, AEs proposed for ASD have to be trained from scratch and do not use prior information or leverage additional training data from other domains. Due to the widespread availability of acoustic datasets and  NNs that were trained on related tasks such as environmental sound classification, \textit{the question arises whether their knowledge, obtained from tasks for which a large amount of data  is available, can be transferred and exploited to increase ASD performance.} This approach is commonly referred to as \textit{Transfer Learning}.\\
Transfer learning is mainly used in computer vision and natural language processing (NLP) but has also started to gain traction in acoustic signal processing.\\
In computer vision, it is common to fine-tune a neural network that was pre-trained on ImageNet for the task of image classification e.g. by only changing some of the last layers and fine-tuning the network on a downstream task. Another approach is to extract some of the activations (e.g. from the penultimate dense layer) of the pretrained network and use them as feature vectors that serve as input to more traditional (e.g. shallow, linear) models. Note that in this case the pretrained network is solely used as feature extractor and its weights are not updated through gradient descent. In both cases, transfer learning considerably reduces the amount of training data needed for the downstream task~\cite{donahue2014decaf}.\\
In NLP, the advent of pretrained language models has lead to substantial improvements on a wide range of tasks~\cite{ruder2018nlpimagenet}. Almost all modern approaches rely on some sort of pretrained (word) representations or networks.
Recent approaches are trained on (masked) language modeling and use transformer~\cite{devlin2018bert} or recurrent neural network architectures~\cite{peters2018deep}. In the first stage the models are trained to solve the very general task of understanding natural language. Afterwards the NNs are fine-tuned or simply used as feature extractors to solve tasks such as sentiment analysis and machine translation.\\
Finally, there exist several pretrained networks from the field of acoustic signal processing.
For example, \cite{chi2020audio} use a transformer based architecture to predict masked Mel-spectrogram frames to obtain contextualized speech representations. \cite{beckmann2019speech} adopt the classic VGG-16 architecture on a spoken word classification task to obtain transferable representations. \cite{Cramer2019} propose a self-supervised audio-visual correspondence task and use the resulting features in conjunction with a simple two layer neural network to achieve state-of-the-art performance for environmental sound classification.\\
Despite of their easy availability, these networks are rarely used in other work i.e. most models are trained from scratch and the possible benefits that the inclusion of pretrained NNs could provide are neglected.\\
In this work, we aim to evaluate the effectiveness of transferring knowledge from related domains using different pretrained NNs for the task of anomalous sound detection. Since no labels are available in ASD, we follow the feature extraction paradigm in conjunction with a Gaussian Mixture Model (GMM). We hypothesize that using the right representation in conjunction with a simple anomaly detection algorithm rivals the performance of the currently widely used autoencoder.
We evaluate our approach with representations from three different domains chosen due to their structural similarity with machine sounds:
(i) Image based representations \textit{(ii)} Environmental sound based representations \textit{(iii)} Music based representations.
From a practical point of view, this approach allows for fast experiments as only shallow models have to be trained and alleviates the burden of designing a suitable neural network architecture and training task.\\
We show that all representations outperform the autoencoder baseline~\cite{Kawaguchi2019} in terms of ASD performance. The best results are obtained with music based representations.
To the best of our knowledge, this is the first study that thoroughly evaluates the efficacy of different feature representations for anomalous sound detection.\\
The rest of the paper is structured as follows:
In Section~\ref{sec:related_work} we briefly review related work followed by a discussion
of our ASD approach in Section~\ref{sec:approach}. Then we go on to describe the experimental setup and the dataset we used to evaluate our approach in Section~\ref{sec:experiments}. The results are discussed in Section~\ref{sec:results}. We close by summarizing our findings and outlining future work in Section~\ref{sec:conclusion}.

\section{\uppercase{Related Work}}
\label{sec:related_work}
\noindent As stated before, the majority of approaches to anomalous sound detection uses deep autoencoders.
\cite{Duman2019} use a convolutional AE to reconstruct Mel-spectrograms in order to detect anomalies in industrial plants. 
\cite{Koizumi2017} concatenate several spectrogram frames as input to a simple, densly connected autoencoder. Here ASD is considered as a statistical hypothesis test where they propose a loss function based on the Neyman-Pearson lemma. However this approach relies on the simulation of anomalous sounds using expensive rejection sampling. \cite{Suefusa2020} use a similar AE architecture to predict the center frame from the remaining frames. This is based upon the observation that the edge frames are usually harder to reconstruct, especially when dealing with non-stationary sounds.
\cite{Marchi2015,bayram2020real} account for the sequential nature of sound and use sequence-to-sequence network architectures to reconstruct auditory spectral features.\\
A slightly different approach is taken by~\cite{Kawaguchi2019}. Here an ensemble of simple autoencoders is used for ASD. Various acoustic front-end algorithms are applied for reverberation and denoising to preprocess the data. This improves performance as the AEs are freed from the burden of reconstructing noise.\\
SNIPER~\cite{koizumi2019sniper} is a novel way to incorporate anomalies that were overlooked during operation without retraining the whole system. For each overlooked anomaly, a new detector is cascaded that is constrained to have a true positive rate (TPR) $=1$ for that specific anomaly. To compute the TPR, a generative model is used to simulate samples of the observed anomaly.\\
Apart from spectrogram based approaches a few ASD approaches that directly operate on the waveform have been developed~\cite{Hayashi2018,Rushe2019}. These methods use causal dilated convolutions~\cite{Oord2016} to predict the next sample, i.e. they are autoregressive models that use the prediction error to measure abnormality.

\section{ASD Approach}
\label{sec:approach}
In this section, we briefly introduce the common acoustic signal processing workflow. Then we describe our ASD appraoch that relies upon pretrained NNs for feature extraction and Gaussian Mixture Models for density estimation in more detail.\\
Most machine learning models do not directly operate on the raw waveform i.e. a sequence  of air pressure measurements. Instead, it is first transformed from the time domain to the frequency domain exploiting the fact that arbitrarily complex waves can be represented as a combinations of simple sinusoids. In practice, this is done using the short-time-fourier transform (STFT). The STFT simply applies the discrete-fourier-transform on small overlapping chunks of the raw signal to account for signals  whose  frequency  characteristics change  over  time.\\
The output of the STFT is a matrix $S \in \mathbb{R}^{F \times T}$ with $F$ frequency bins and $T$ time frames and is called the spectrogram. Each cell represent the amplitude (the normalized squared magnitude) of the corresponding frequency bin at some point in time.\\
To account for the fact that the human perception of pitch is logarithmic in nature, i.e. humans are more discriminative at lower frequencies and less discriminative at higher frequencies, one additionally transforms the frequency bins into the Mel-scale using the Mel-filterbank that is composed of overlapping triangular filters. The Mel-scale is an empirically defined perceptual scale that mirrors human perception. Filters are small for low frequencies and are of increasing width for higher frequencies. They aggregate the energy of consecutive frequency bins to describe how much energy exists in various frequency regions.
This can be seen as a re-binning procedure that alters the number of bins from $F$ to $\mathcal{M}$, $\mathcal{M} < F$ where $\mathcal{M}$ is the number of Mel-filters used. Finally the log of the energies is taken to convert from amplitude to decibel (dB).
The resulting Mel-spectrogram is a compact visual representation of audio that can be used as input to a machine-learning pipeline. Moreover, the Mel-spectrogram can be treated as an image of the underlying signal and one can therefore resort to computer vision approaches such as convolutional neural networks (CNNs).\\
We assume that the ASD system monitors the entity for an application specific, predefined period of time (decision horizon) until the degree of abnormality (or normality) needs to be estimated and that we are given a dataset of normal operation recordings only. Therefore, we assume that the dataset $\mathcal{D}$ is a set of $n$ (equal-length) Mel-spectrograms 
\begin{equation}
    \label{eq:dataset}
    \mathcal{D}= M_1, \dots, M_n \in \mathbb{R}^{\mathcal{M} \times T}
\end{equation}
where $T$ is proportional to the decision horizon.\\
It is common to subdivide the computation of the final anomaly score into a sequence of smaller predictions that operate on a smaller timescale. For example, the decision horizon might be $10$ seconds but the ASD system outputs an anomaly score every $0.5$ seconds. This is a simple strategy to avoid overlooking subtle short-term anomalies and causes a trade-of between local and global structure. 
A score for the whole decision horizon is obtained by aggregation. 
In our case, we transform each Mel-spectrogram $M_i,\, i = 1 \dots n$ into a sequence of feature representations $V_i \in \mathbb{R}^{D \times \Tilde{T}}$ by using a sliding window across the columns (time dimension). For each window of $t$ frames, a feature extractor
\begin{equation}
     \Phi: \mathbb{R}^{\mathcal{M} \times t} \rightarrow \mathbb{R}^{D}
\end{equation}
extracts a $D$ dimensional feature vector that is more semantically compact than raw Mel-spectrogram features. Furthermore, the feature extractor leverages prior knowledge obtained through its pretext task. Then the window is moved $h$ frames to the right. Consequently, the transformed dataset $\mathcal{D}_{\Phi}$ has the form 
\begin{equation}
    \label{eq:dataset_transformed}
    \mathcal{D}_{\Phi}= V_1, \dots, V_n \in \mathbb{R}^{D \times \Tilde{T}}
\end{equation}
The number of feature vectors $\Tilde{T}$ is given by $\frac{T-n}{h}$. Note that if $\frac{T-n}{h} \not\in \mathbb{N}$ an appropriate padding strategy has to be applied.\\
Since no labels are available we propose to fit a Gaussian mixture model (GMM) with $K$ components to individual feature vectors in order to being able to compute their density. The density can in turn be used to describe the degree of normality of a single feature vector.\\
The GMM was chosen as anomaly detector as it was shown~\cite{Muller2020a} to outperform other models on the task of anomalous sound detection when using image based features. Moreover, GMMs are fast and easy to train due to the ready availability of reliable implementations, have a sound probabilistic interpretation and have the capability to model arbitrary densities.\\
The density of a single feature vector $x \in \mathbb{R}^D$ is given by:
\begin{equation}
    \label{eq:gmm}
    p(x) = \sum_{k=1}^{K} \lambda_i * N(x|\mu_i, \Sigma_i)
\end{equation}
$\mu_k$, $\Sigma_k$ and $\lambda_k$ are the mean, covariance matrix and weight of the $k$th mixture component where $\sum_{k=1}^{K}\lambda_k = 1$.
We compute the anomaly score for some $V_i$ as follows:
\begin{equation}
    \label{eq:score}
    p(V_i) = p(V_{i|*,1}, \dots, V_{i|*,\Tilde{T}}) =\prod_{j=1}^{\Tilde{T}} p(V_{i|*,j})
\end{equation}
It is important to emphasize that Equations~\ref{eq:gmm} and~\ref{eq:score} are highly dependent on expressive features that enable learning a meaningful notion of normality. Note that in Equation~\ref{eq:score} we have assumed independence between consecutive feature vectors for simplicity and used $V_{i|*,j}$ to select the $j$th column vector from $V_i$.

\section{\uppercase{Experiments}}
\label{sec:experiments}
\noindent In this section we first introduce the dataset that was used to evaluate our hypothesis. Then we describe each deployed feature representation in more detail, followed by the experimental setup and a brief discussion of how the GMM's hyperparameters were chosen.
\subsection{Dataset}
\begin{figure}
  \centering
  \includegraphics[width=0.95\linewidth]{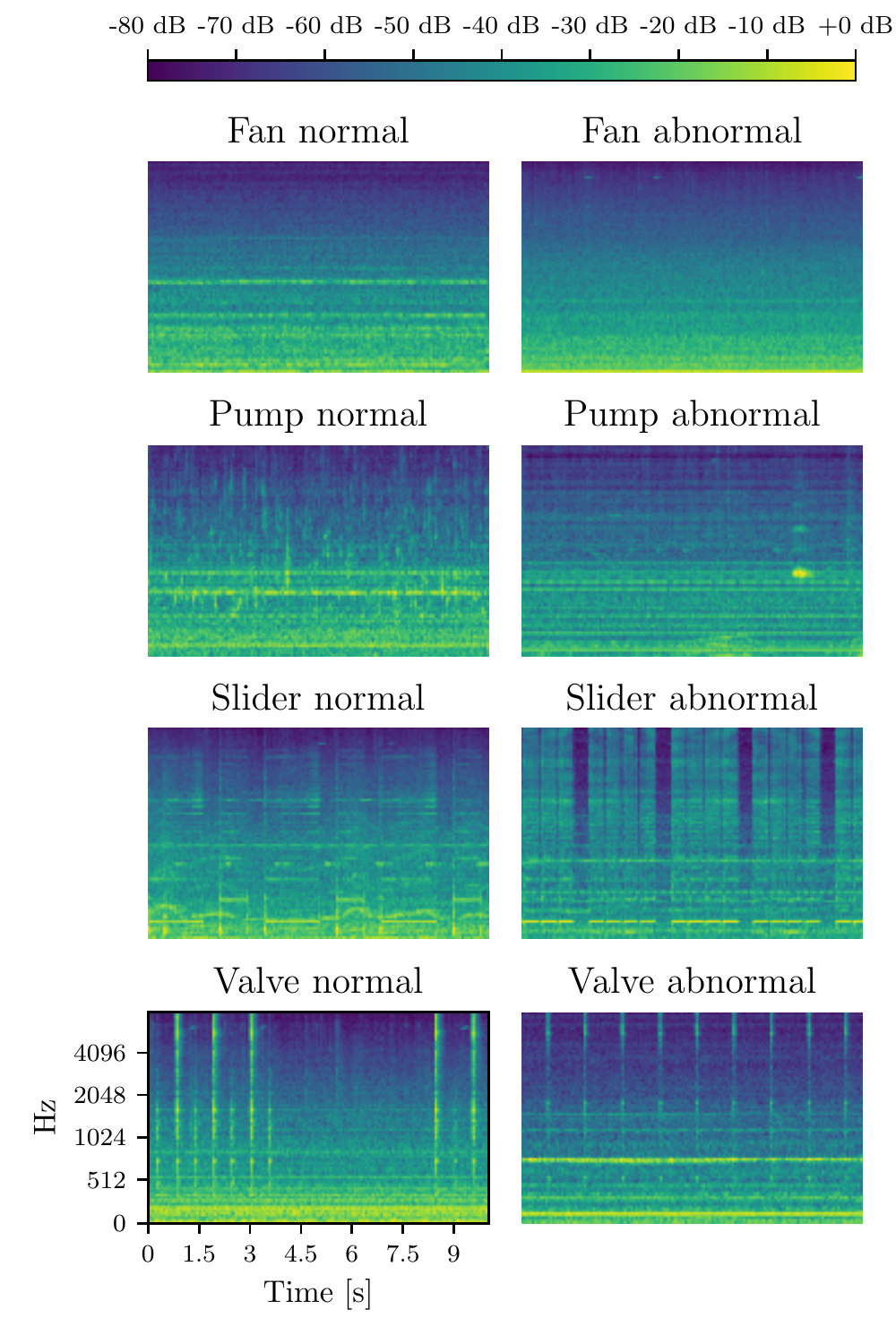}
  \caption{Exemplary Mel-spectrograms of normal and abnormal operation with SNR$=6$dB and machine id $=0$.}
    \label{fig:mels}
\end{figure}

To study the efficacy of different feature representations for anomalous sound detection, we use the recently published MIMII dataset~\cite{Purohit2019}. It consists of recordings from four different machine types: fans, pumps, slide rails and valves under normal and abnormal operation. Examples for anomalous conditions are: leakage, clogging, voltage change, a loose belt, rail damage or no grease. Additionally. For each machine type there are sounds from four different machine models (ID $0,2,4$ and $6$). Lastly, there exist three different versions of each recording where real-world background-noises from a factory are mixed with the machine sound according to a signal-to-noise ratio (SNR) of $6$dB, $0$dB and $-6$dB. Recordings with a SNR of $-6$dB are the most challenging as they are highly contaminated with background noise.
A practically applicable and reliable anomalous sound detection system should be able to yield good results across all combinations. In total, there are $26092$ normal condition and $6065$ anomalous condition segments which are divided over $4*4*3= 48$ (machine types * machine ids * SNRs) different datasets. All recordings are sampled at $16$kHz and have a duration of $10$ seconds. See Figure~\ref{fig:mels} for some exemplary Mel-spectrograms.\\

\subsection{Feature Representations}
We evaluate feature representations for ASD from three different domains: music, images and environmental sounds.\\

The following NNs were trained on data from an acoustically highly similar domain to the one studied in this work namely, environmental sounds:\\
\textbf{VGGish}~\cite{Hershey2017}\footnote{\texttt{https://github.com/beasteers/VGGish}}\\
    VGGish is a variant of the VGG architecture that has successfully been applied to image classification.
    Compared to VGG16, the number of weight layers is reduced to $11$. It consists of four groups of convolution+maxpooling followed by a single $128$ dimensional fully connected layer. It was trained to classify the soundtracks of a dataset of $\approx$ 70M training videos.\\
\textbf{L3env}~\cite{Cramer2019}\footnote{\texttt{https://openl3.readthedocs.io/}}\\
        Unlike the other networks used for feature extraction in this work that were trained on a supervised learning task, L3 was trained in a self-supervised manner. 
         The audio-visual correspondence task, i.e. predicting whether a video frame corresponds to an audio frame, removes the need for any external labels. Training was carried out using $\approx$ 195K Videos of natural acoustic environments (e.g. human and animal sounds) extracted from AudioSet.\\

Musically related tasks such as music-tagging rely upon a rich set of features such as tone, timbre and pitch.
We assume that such features are also important for ASD. The two NNs below were trained on musically related tasks:\\
\textbf{MusiCNN}~\cite{Pons2019}\footnote{\texttt{https://github.com/jordipons/musicnn}}\\
This CNN based architecture uses a musically motivated front-end, a densely connected mid-end and a temporal-pooling back-end. The shapes of the CNN filters are explicitly designed to account for phenomena and concepts that typically appear in music (e.g. pitch, timbre, tempo). Since the filter dimensions on spectrograms comprise of time and frequency, wider filters are used to account for longer temporal dependencies and higher filters are used to capture timbral features.
The MusiCNN we use in this work, was trained totag music on the MagnaTagATune dataset. Moreover, we choose the activations from the mean-pooling layer as feature representation.\\
\textbf{L3music}~\cite{Cramer2019}\footnote{\texttt{https://openl3.readthedocs.io/}}\\
The only difference to L3env is, that L3music was trained on $\approx$ 296K videos of people playing musical instrument extracted from AudioSet.\\

Lastly, the following NNs were trained on the task of image classification on ImageNet. The Mel-spectrograms are converted to $224\times224$ images using the \textit{Viridis} colormap\footnote{Note that we also evaluated different color-maps but have found the differences in the results to be neglectable.} as suggested by \cite{Amiriparian2017} in the context of snore-sound classification. Then the RGB values are standardized using the values obtained from ImageNet. We chose images because we observed that anomalous sound patterns can often be spotted visually in their spectral representation~\cite{Muller2020a}.\\
\textbf{ResNet34}~\cite{He2016}\\
ResNet uses skip-connections to give the network the ability to bypass blocks of convolutions.
This is done to ease the vanishing gradient problem that would occur due to an increased network depth. Depth allows the network to extract a richer, more informative features. We use the activations from the penultimate layer (right before the 1K-way classification head) as representation.\\
\textbf{DenseNet121}~\cite{Huang2018}\\
Just as ResNet, DenseNet was designed to enable greater network depth while being more parameter efficient. Here, the next convolutional layer receives the feature maps from all previous layers and computes a small number of feature maps to add to the stack according to a growth rate. This process is interleaved with $1\times1$ convolutions and average pooling to reduce the number of channels and the size of the feature maps, respectively.
We average-pool the activations from the penultimate layer to receive a flat feature representation.\\

To compare the representations above with the more common AE-based approach we implemented the following AE:\\
\textbf{Autoencoder}(AE)~\cite{Purohit2019}\\
    This autoencoder was proposed by the authors of the MIMII dataset and serves as a strong baseline to contrast transfer learning approaches with the more commonly applied method of reconstruction error based anomaly detection. Every $5$ columns (time dimension) of the Mel-spectrogram are concatenated to form a feature vector $\in \mathbb{R}^{320}$
    which serves as input to a dense AE architecture (same as~\cite{Purohit2019}). The mean squared reconstruction error is used as loss function and as anomaly score. The AEs are trained with a batch size of $128$ for $50$ epochs, a learning rate of $0.002$ and a L2 regularization of $10^{-5}$.\\
    
A brief overview of the different feature extractors and their corresponding training domain and representation representation size (dimensionality) is given in Table~\ref{tab:extractors1}.

\subsection{Experimental Setup}
\label{sec:experiments_steup}
\begin{figure*}
  \centering
  \includegraphics[width=0.98\textwidth]{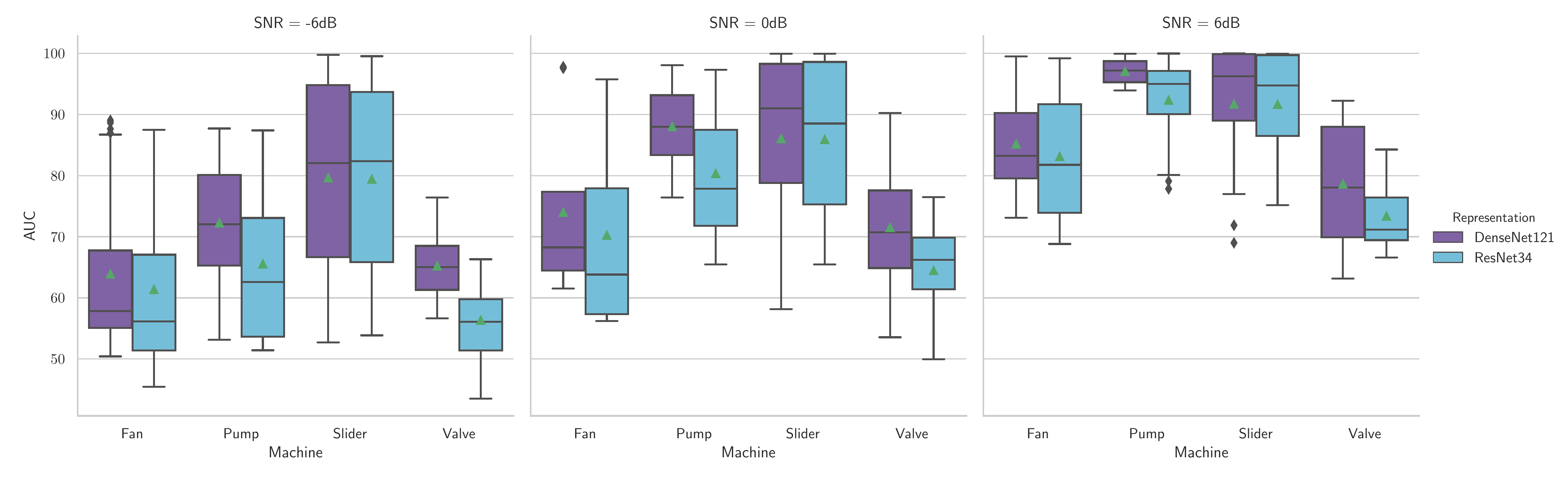}
  \caption{Comparison between \textbf{image} based representations.}  
   \label{fig:imgs}
\end{figure*}

\begin{figure*}
  \centering
  \includegraphics[width=.98\linewidth]{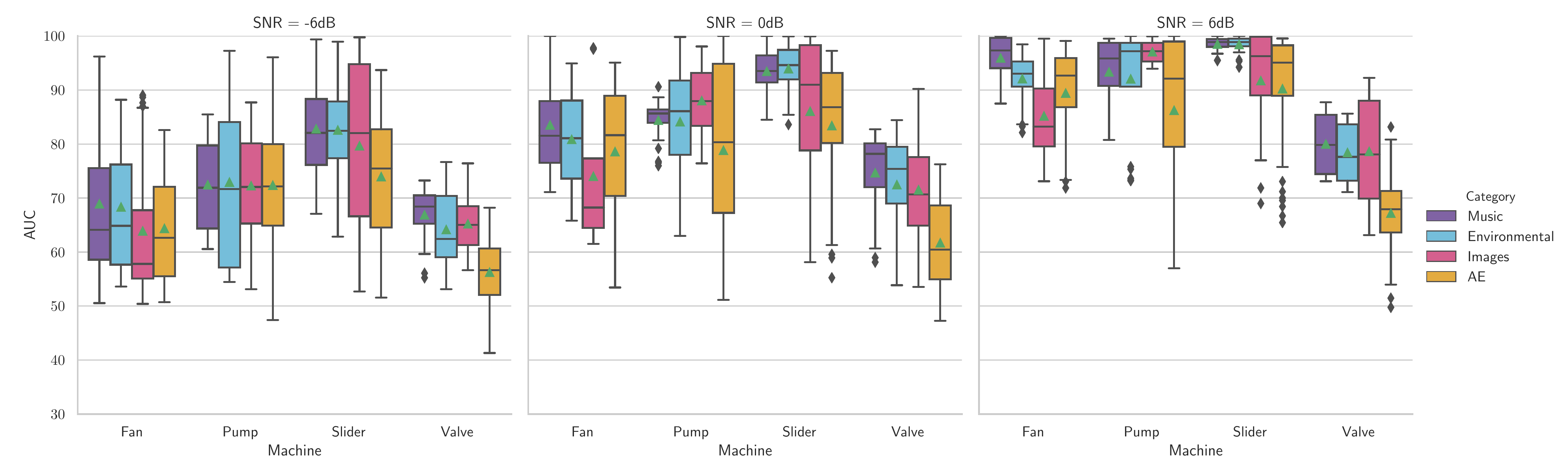}
  \caption{Comparison between the \textbf{best performing} representations and the autoencoder baseline. Music representations: MusiCNN for Fan and Pump, L3music for Slider and Valve. Environmental Sounds: L3env. Images: DenseNet121.}
  \label{fig:comparison}
\end{figure*}

To study the efficacy of the different feature extractors, we propose the following workflow:
\begin{enumerate}
    \item Compute the feature representations for each $10$s long sound sample in the train set with a window size of $1$s and an overlap of $0.5$s. This results in $20$ feature vectors per sample.
    \item Apply normalization to the resulting representations and use Principal Components Analysis (PCA) to reduce their dimensionality such that $98$\% of the representations variance is retained.
    \item Fit a Gaussian Mixture Model (GMM) to the transformed data. 
    \item At test time, obtain the transformed feature representations for each $10$s long sound sample (Steps $1$ \& $2$) in the test set. The anomaly score for a single transformed feature representation is given by the weighted negative log probabilities. By mean-pooling the scores over each sound sample, one obtains the final anomaly scores.
\end{enumerate}
A high-level overview of this process is depicted in Figure~\ref{fig:workflow}.\\
Model performance is measured with the \textit{Area Under the Receiver Operating Characteristics} (AUC) which quantifies how well a model can distinguish between normal and anomalous operation across all possible thresholds.\\
For each combination of machine type, machine id, SNR and feature representation a separate GMM is trained.\\ To form the test set, the same amount of normal operation data is randomly removed from the train set as there is anomalous data, i.e. training is done on the remaining  normal operation data, anomalous data is never seen during training and the test set is balanced. Each experiment is repeated five times across five different seeds.\\
First, the results are conditioned on the representation, the SNR and the machine type to obtain a corresponding performance distributions. We use Box-plots to report these distributions grouped by the domain of the representations. The results for image, music and environmental sound based representations are depicted in Figure~\ref{fig:imgs}, Figure~\ref{fig:music} and Figure~\ref{fig:env}, respectively. Then we select the best performing (feature extractor - machine type) pairs and compare the results by domain as well as with the autoencoder baseline in Figure~\ref{fig:comparison}.\\
In Table~\ref{tab:big_results} we additionally condition the results from Figure~\ref{fig:comparison} on the machine id to gain more insights on how the models perform on each individual machine id. Each cell represents the average AUC $\pm$ one standard deviation.\\
A detailed discussion of the results follows in Section~\ref{sec:results}.
\begingroup
\setlength{\tabcolsep}{6pt} 
\renewcommand{\arraystretch}{1.1} 
\begin{table}[]
\centering
\caption{Overview of the different feature extractors.}
\label{tab:extractors1}
\begin{tabular}{lcc}
\textbf{Name} & \textbf{Trained on} & \textbf{Rep. size} \\ \hline
\multicolumn{1}{l|}{L3env} & \multirow{2}{*}{\begin{tabular}[c]{@{}c@{}}Environmental\\ sounds\end{tabular}} & 512 \\
\multicolumn{1}{l|}{VGGish} &  & 128 \\ \hline
\multicolumn{1}{l|}{ResNet34} & \multirow{2}{*}{Images} & 512 \\
\multicolumn{1}{l|}{DenseNet121} &  & 1024 \\ \hline
\multicolumn{1}{l|}{MusiCNN} & \multirow{2}{*}{Music} & 753 \\
\multicolumn{1}{l|}{L3music} &  & 512 \\ \hline
\end{tabular}
\end{table}
\endgroup


\subsection{Choice of Hyperparameters}
\begin{figure}
  \centering
  \includegraphics[width=\linewidth]{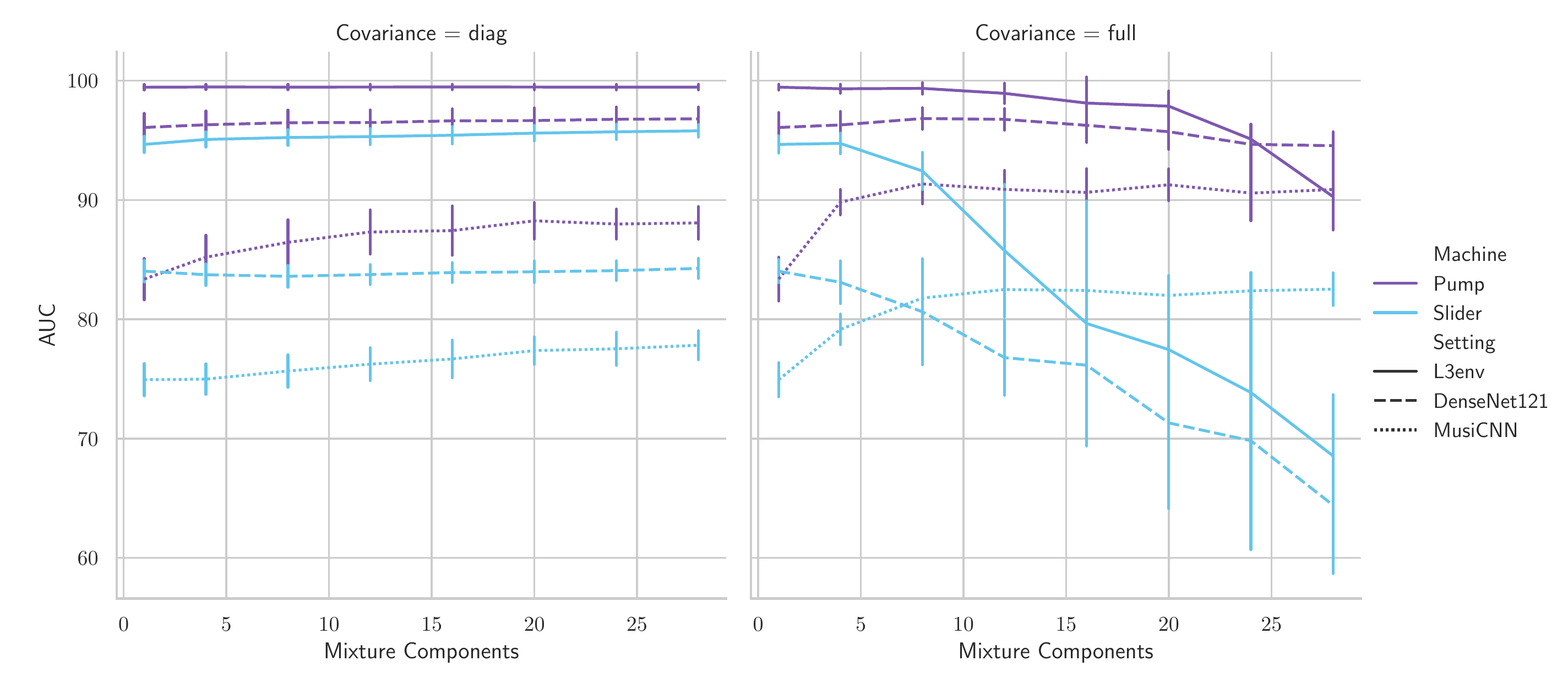}
  \caption{Small scale study to determine the GMM hyperparameters.}
  \label{fig:gmm}
\end{figure}

The crucial hyperparameters of GMMs are the number of mixtures and the covariance matrix type.
To determine suitable hyperparameters, we randomly selected three feature extractors and two machine types with SNR$=0$dB and ID$=0$. Then we computed the average AUC with a varying number of mixture components $\{1,4,8,\dots 28\}$ with full and diagonal covariance matrices (Figure \ref{fig:gmm}).\\
While the features extracted from MusiCNN appear to be stable with respect to the number of mixtures and the covariance type, results for other representations quickly decline when using full covariance matrices and an increasing number of mixtures. Hence, the GMM becomes too expressive. This effect is not observed when using diagonal covariance matrices as the results are stable and slightly increase with the number of mixtures.\\
Since anomalous data is usually scarce, it is not feasible to tune the hyperparameters on every setting and consequently a stable set of hyperparameters is desirable. Therefore, we use a GMM with $20$ mixture components and diagonal covariance matrices for all experiments in this work.

\section{\uppercase{Results}}
\label{sec:results}
\noindent In this section we first discuss the results for each individual feature extractor domain. Then we present findings that apply to all domains. Finally, we discuss the limitations of our analysis.
\begingroup
\setlength{\tabcolsep}{2.4pt} 
\renewcommand{\arraystretch}{1.0} 
\begin{table*}[h]
\centering
\scriptsize
\caption{Mean AUCs for all experiments.}
\label{tab:big_results}
\begin{tabular}{cc|cccc|cccc|cccc}
\hline
\multicolumn{2}{c|}{SNR} & \multicolumn{4}{c|}{-6dB} & \multicolumn{4}{c|}{0dB} & \multicolumn{4}{c}{6dB} \\
Machine & ID & \textbf{AE} & \textbf{Env.} & \textbf{Imgs.} & \textbf{Music} & \textbf{AE} & \textbf{Env.} & \textbf{Imgs.} & \textbf{Music} & \textbf{AE} & \textbf{Env.} & \textbf{Imgs.} & \textbf{Music} \\ \hline
\multirow{4}{*}{\textbf{Fan}} & 0 & 53.0$\pm$1.6 & 58.7$\pm$1.2 & 57.3$\pm$1.3 & 61.4$\pm$1.0 & 58.0$\pm$3.5 & 67.0$\pm$0.8 & 62.7$\pm$0.8 & 71.6$\pm$0.4 & 74.5$\pm$1.7 & 83.7$\pm$1.3 & 74.1$\pm$0.8 & 88.5$\pm$0.6 \\
 & 2 & 68.7$\pm$1.6 & 70.9$\pm$1.1 & 58.5$\pm$1.7 & 66.6$\pm$1.7 & 85.9$\pm$1.5 & 85.1$\pm$0.6 & 69.7$\pm$0.9 & 82.9$\pm$0.9 & 93.0$\pm$1.3 & 92.8$\pm$0.5 & 81.4$\pm$0.5 & 96.4$\pm$0.5 \\
 & 4 & 56.8$\pm$1.2 & 56.3$\pm$1.8 & 52.0$\pm$1.1 & 52.8$\pm$1.5 & 76.5$\pm$2.1 & 76.5$\pm$0.7 & 65.9$\pm$1.4 & 79.6$\pm$1.4 & 91.5$\pm$1.2 & 93.5$\pm$0.6 & 85.9$\pm$1.1 & 98.9$\pm$0.7 \\
 & 6 & 78.9$\pm$2.1 & 87.4$\pm$0.6 & 87.8$\pm$1.0 & 94.8$\pm$1.0 & 93.9$\pm$0.8 & 94.7$\pm$0.2 & 97.7$\pm$0.1 & 99.9$\pm$0.1 & 98.7$\pm$0.3 & 98.4$\pm$0.2 & 99.4$\pm$0.1 & 100.0$\pm$0 \\ \cline{2-14} 
 & $\varnothing$ & \multicolumn{1}{l}{64.4$\pm$10.4} & \multicolumn{1}{l}{68.3$\pm$12.7} & \multicolumn{1}{l}{63.9$\pm$14.4} & \multicolumn{1}{l|}{\textbf{68.9$\pm$16.2}} & \multicolumn{1}{l}{78.6$\pm$13.7} & \multicolumn{1}{l}{80.8$\pm$10.5} & \multicolumn{1}{l}{74.0$\pm$14.3} & \multicolumn{1}{l|}{\textbf{83.5$\pm$10.6}} & \multicolumn{1}{l}{89.4$\pm$9.2} & \multicolumn{1}{l}{92.1$\pm$5.5} & \multicolumn{1}{l}{85.2$\pm$9.5} & \multicolumn{1}{l}{\textbf{95.9$\pm$4.6}} \\ \hline
\multirow{4}{*}{\textbf{Pump}} & 0 & 71.4$\pm$2.8 & 78.6$\pm$1.5 & 75.8$\pm$2.1 & 84.0$\pm$1.5 & 72.5$\pm$1.7 & 87.9$\pm$1.2 & 85.1$\pm$1.1 & 85.4$\pm$0.4 & 87.1$\pm$1.0 & 96.0$\pm$0.5 & 94.5$\pm$0.6 & 94.1$\pm$0.4 \\
 & 2 & 53.2$\pm$3.0 & 55.4$\pm$1.1 & 68.1$\pm$2.1 & 62.3$\pm$1.3 & 55.1$\pm$2.4 & 65.9$\pm$2.5 & 90.4$\pm$1.3 & 77.8$\pm$2.0 & 60.1$\pm$1.5 & 74.3$\pm$1.2 & 97.6$\pm$0.4 & 81.6$\pm$0.9 \\
 & 4 & 93.0$\pm$2.3 & 96.1$\pm$1.3 & 86.5$\pm$1.1 & 76.2$\pm$2.4 & 99.6$\pm$0.6 & 99.5$\pm$0.3 & 96.7$\pm$1.1 & 88.3$\pm$1.6 & 100.0$\pm$0 & 100.0$\pm$0. & 99.8$\pm$0.1 & 98.9$\pm$1.0 \\
 & 6 & 71.8$\pm$3.0 & 61.6$\pm$3.2 & 58.7$\pm$3.8 & 67.2$\pm$2.6 & 88.1$\pm$2.6 & 83.1$\pm$1.8 & 79.9$\pm$2.4 & 86.0$\pm$0.7 & 97.8$\pm$1.0 & 97.9$\pm$0.3 & 96.4$\pm$1.6 & 98.7$\pm$0.1 \\ \cline{2-14} 
 & $\varnothing$ & \multicolumn{1}{l}{72.4$\pm$14.5} & \multicolumn{1}{l}{\textbf{72.9$\pm$16.4}} & \multicolumn{1}{l}{72.3$\pm$10.7} & \multicolumn{1}{l|}{72.4$\pm$8.8} & \multicolumn{1}{l}{78.8$\pm$17.0} & \multicolumn{1}{l}{84.1$\pm$12.5} & \multicolumn{1}{l}{\textbf{88.0$\pm$6.5}} & \multicolumn{1}{l|}{84.4$\pm$4.2} & \multicolumn{1}{l}{86.2$\pm$16.1} & \multicolumn{1}{l}{92.0$\pm$10.6} & \multicolumn{1}{l}{\textbf{97.1$\pm$2.2}} & \multicolumn{1}{l}{93.3$\pm$7.2} \\ \hline
\multirow{4}{*}{\textbf{Slider}} & 0 & 91.5$\pm$1.9 & 98.5$\pm$0.3 & 99.5$\pm$0.3 & 99.1$\pm$0.3 & 96.4$\pm$0.8 & 99.8$\pm$0.1 & 99.9$\pm$0.1 & 99.9$\pm$0.1 & 99.2$\pm$0.3 & 100.0$\pm$0 & 100.0$\pm$0 & 100.0$\pm$0 \\
 & 2 & 77.9$\pm$1.6 & 83.0$\pm$0.9 & 91.7$\pm$1.0 & 83.2$\pm$1.0 & 86.5$\pm$0.6 & 93.1$\pm$0.8 & 97.0$\pm$0.5 & 92.7$\pm$0.6 & 94.4$\pm$0.8 & 99.0$\pm$0.3 & 99.5$\pm$0.2 & 98.9$\pm$0.2 \\
 & 4 & 71.0$\pm$2.7 & 82.0$\pm$1.4 & 70.6$\pm$2.3 & 79.2$\pm$1.7 & 88.9$\pm$3.0 & 95.6$\pm$0.7 & 84.0$\pm$1.0 & 94.4$\pm$0.8 & 95.6$\pm$1.8 & 98.8$\pm$0.3 & 92.4$\pm$0.5 & 98.6$\pm$0.4 \\
 & 6 & 55.3$\pm$2.8 & 66.8$\pm$2.4 & 56.8$\pm$3.8 & 69.8$\pm$1.8 & 61.8$\pm$3.1 & 87.3$\pm$2.8 & 63.3$\pm$3.1 & 86.7$\pm$2.2 & 71.7$\pm$4.6 & 95.9$\pm$1.3 & 74.9$\pm$4.4 & 96.4$\pm$0.9 \\ \cline{2-14} 
 & $\varnothing$ & \multicolumn{1}{l}{73.9$\pm$13.4} & \multicolumn{1}{l}{82.6$\pm$11.6} & \multicolumn{1}{l}{79.7$\pm$17.5} & \multicolumn{1}{l|}{\textbf{82.8$\pm$10.9}} & \multicolumn{1}{l}{83.4$\pm$13.3} & \multicolumn{1}{l}{\textbf{93.9$\pm$4.8}} & \multicolumn{1}{l}{86.1$\pm$14.9} & \multicolumn{1}{l|}{93.5$\pm$5.0} & \multicolumn{1}{l}{90.2$\pm$11.3} & \multicolumn{1}{l}{98.4$\pm$1.7} & \multicolumn{1}{l}{91.7$\pm$10.6} & \multicolumn{1}{l}{\textbf{98.5$\pm$1.4}} \\ \hline
\multirow{4}{*}{\textbf{Valve}} & 0 & 50.3$\pm$4.7 & 61.0$\pm$1.7 & 73.4$\pm$1.7 & 70.3$\pm$2.2 & 51.4$\pm$2.4 & 75.4$\pm$3.6 & 88.6$\pm$1.7 & 78.1$\pm$2.9 & 60.5$\pm$8.7 & 74.2$\pm$0.7 & 90.5$\pm$1.7 & 75.2$\pm$1.2 \\
 & 2 & 62.0$\pm$3.5 & 74.9$\pm$1.9 & 63.9$\pm$3.0 & 70.3$\pm$2.4 & 72.2$\pm$2.9 & 82.0$\pm$2.1 & 70.8$\pm$1.7 & 80.2$\pm$1.6 & 75.0$\pm$6.0 & 85.0$\pm$1.0 & 72.6$\pm$1.5 & 86.3$\pm$0.8 \\
 & 4 & 61.5$\pm$3.0 & 64.8$\pm$3.3 & 64.2$\pm$2.0 & 68.2$\pm$1.7 & 65.6$\pm$4.2 & 74.6$\pm$1.9 & 70.8$\pm$2.7 & 79.6$\pm$2.6 & 66.9$\pm$3.4 & 82.4$\pm$1.8 & 85.4$\pm$2.5 & 84.2$\pm$0.8 \\
 & 6 & 51.2$\pm$2.9 & 55.8$\pm$2.6 & 59.4$\pm$3.3 & 58.7$\pm$2.9 & 57.5$\pm$2.0 & 57.8$\pm$3.6 & 55.9$\pm$1.7 & 60.7$\pm$2.6 & 66.3$\pm$5.0 & 72.0$\pm$0.8 & 66.0$\pm$2.1 & 74.2$\pm$1.1 \\ \cline{2-14} 
 & $\varnothing$ & \multicolumn{1}{l}{56.2$\pm$6.6} & \multicolumn{1}{l}{64.2$\pm$7.5} & \multicolumn{1}{l}{65.2$\pm$5.8} & \multicolumn{1}{l|}{\textbf{66.9$\pm$5.4}} & \multicolumn{1}{l}{61.7$\pm$8.5} & \multicolumn{1}{l}{72.5$\pm$9.6} & \multicolumn{1}{l}{71.5$\pm$12.0} & \multicolumn{1}{l|}{\textbf{74.7$\pm$8.6}} & \multicolumn{1}{l}{67.2$\pm$7.8} & \multicolumn{1}{l}{78.4$\pm$5.7} & \multicolumn{1}{l}{78.6$\pm$10.2} & \multicolumn{1}{l}{\textbf{80.0$\pm$5.5}} \\ \hline
\end{tabular}
\end{table*}
\endgroup

\subsection{Image based Representations}
DenseNet121 outperforms ResNet34 in terms of the mean and median AUC on all machine types except for slider where the difference is insignificant. Moreover, there are no major differences with respect to the variance. The most striking performance differences are observed on pump and valve.\\
A possible explanation for the superiority of DenseNet121 is that it also outperforms ResNet34 in terms of classification accuracy on ImageNet by extracting a richer, more nuanced and more discriminative set of features and that this superiority also carries over to our setting.\\
Recent work~\cite{Muller2020a} did not evaluate their method on DenseNet based representations and concluded that ResNet based features are best suited for anomalous sound detection. In contrast, here we have found that in our setting, ResNet is outperformed by DenseNet.
Our results indicate that further studies are necessary to explore the space of image based feature extractors for ASD.

\subsection{Music based Representations}
\label{sec:music_results}
\begin{figure*}
  \centering
  \includegraphics[width=.98\textwidth]{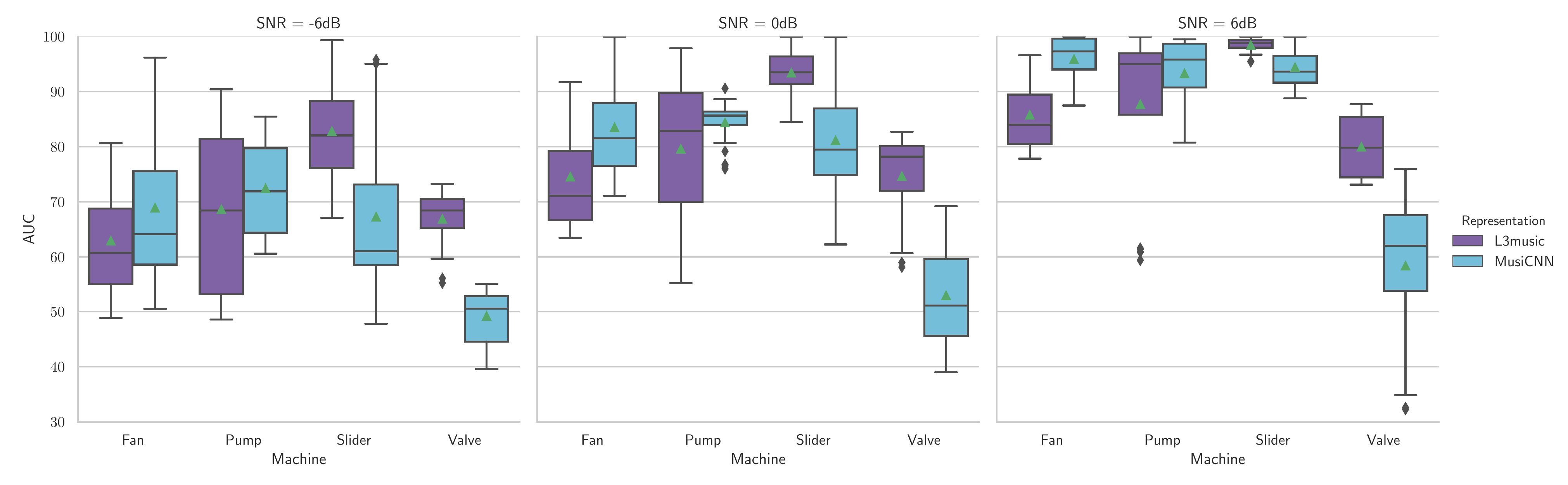}
  \caption{Comparison between \textbf{music} based representations.}
  \label{fig:music}
\end{figure*}

From Figure~\ref{fig:music} we can observe that L3music shows significantly better results on slider and valve than MusiCNN. On the other hand, MusiCNN outperforms L3music on fan and pump.\\
While fan and pump have a stationary sound pattern, slider and valve exhibit non-stationary patterns. Our interpretation is that MusiCNN's rectangular horizontal filters (time dimension) are especially well suited to extract features that describe the normal operation of stationary sounds since they are almost constant and vary only slightly over time. Then, anomalies will be characterized by the absence of certain features for example because the machine randomly stops operating, suddenly exhibits irregular sounds or the pitch of the sound changes due to damages. For slider and valve, MusiCNN's music tagging task might be too restrictive compared to the very general audio-visual correspondence task of L3music.

\subsection{Environmental Sound based Representations}
\begin{figure*}
  \centering
  \includegraphics[width=.98\linewidth]{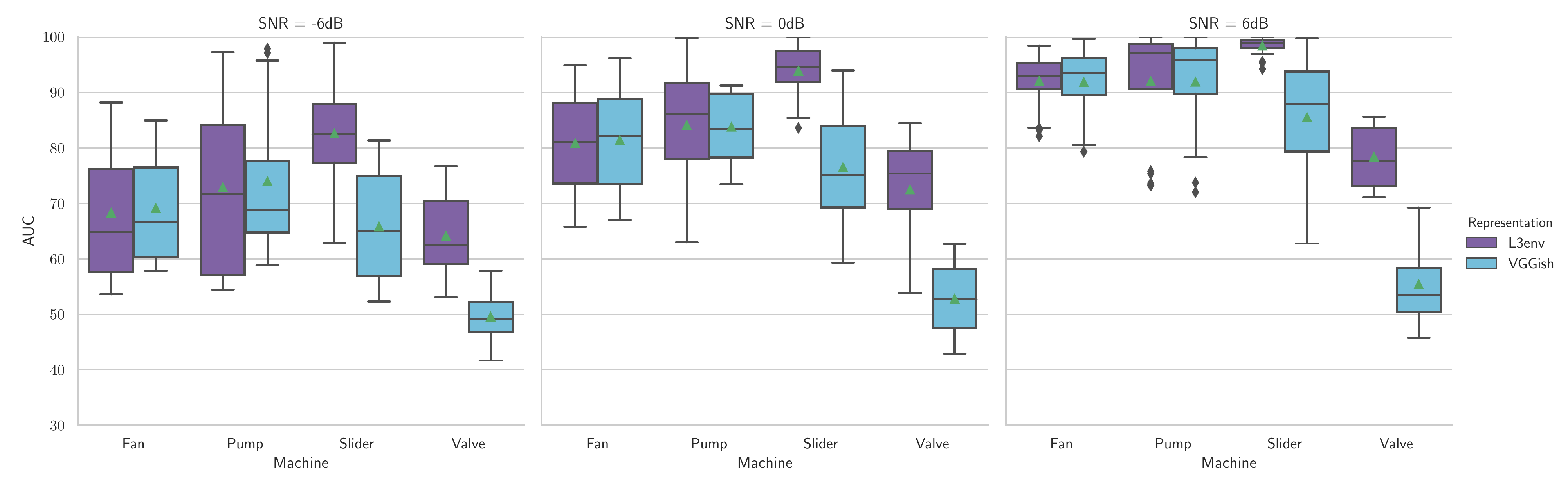}
  \caption{Comparison between \textbf{environmental sound} based representations.}
  \label{fig:env}
\end{figure*}

From the inspection of Figure~\ref{fig:env} we can conclude that  L3env yields far better results on slider and valve compared to VGGish. On fan and pump, both feature extractors perform on-par.\\
Just as in Section~\ref{sec:music_results}, the most obvious performance differences can be observed on non-stationary machine sounds.
We noticed that on slider and valve, the VGGish feature representations are highly redundant. That is, PCA reduces representations down to a very small number of features. Essentially, VGGish extracts very similar features for all recordings thereby hindering the creation of an informative normal operation model. \\
Previous work \cite{Cramer2019} used L3 representations for sound event detection and reports superior performance compared to VGGish. In our case these results also hold for ASD. Moreover, we conjuncture that the self-supervised audio-visual correspondence task extracts a richer set of features as it is a more complex task that needs more expressive, general purpose features compared to VGGish's more narrow classification objective.
\subsection{General Observations}
Generally, the introduction of background noise considerably reduces the performance of all representations. Moreover, all representations are affected by background noise in the same way i.e. noise does not change the performance ranking.\\
Fan is the machine type that suffers the most from increased background noise. This is because fan sounds and background noise exhibit a similar sound pattern, making them harder to distinguish.\\
We found the valve machine type samples (highly non-stationary) to be the hardest to detect anomalous operations on, and slider to be the easiest (clearly visible anomalies in the Mel-spectrograms).\\
When averaged over the machine id  (Table~\ref{tab:big_results}), all transfer learning approaches outperform the AE baseline. Thus, this confirms our hypothesis that transferring knowledge from pretrained NNs increases ASD performance.\\
Music based representations achieve the best results with $8/12$ top performances. Both image and environmental sound based representations account for $2/12$ top performances, each.\\
These results challenge the intuition that closely matching the domain results in better downstream task performance as one might have suspected environmental sound based representations to yield the best results because they have the smallest domain mismatch with machine sounds.
However, it might be more important to use audio content that maximizes the discriminative power of the representations, independently of the downstream domain~\cite{Cramer2019}. In the case of L3 based representations, people playing musical instruments have a greater degree of audio visual correspondence than environmental videos. Moreover, music itself provides a richer, more diverse training signal than environmental sounds.

\subsection{Limitations of the Analysis}
Evaluating the different representations presents a challenge as it is hard to find the particular reasons why one representation works better than another since the feature extractors were trained on different dataset with a varying amount of data. Additionally, they all use different (black-box) neural network architectures and slightly different Mel-spectrogram parameters. Moreover, their domain varies, the training task was either supervised or self-supervised and no information on the specific type of anomaly of a recording is provided by the MIMII dataset.

\section{\uppercase{Conclusion}}
\label{sec:conclusion}
\noindent In this work, we evaluated the effectiveness of transferring knowledge from pretrained NNs for anomalous sound detection using the feature extraction paradigm.
Our approach was evaluated with feature representations from the image, environmental sound and music domain and allows for fast experimentation as only shallow models have to be trained.
We showed that almost all representations yield competitive ASD performance with an advantage for music based representations. Thus, we have found that even under a domain mismatch between the feature extractor and the downstream ASD task, the studied representations are suitable for ASD.
This suggests that pretraining on a closely related domain might not always be necessary which results in a greater flexibility in the choice of pretraining strategies and datasets.
\textit{The key finding of this work is that a relatively simple experimental setup based on transfer learning can yield competitive ASD performance without the need to develop a completely new model.} Hence, we argue that future approaches should compare themselves against transfer learning approaches. Furthermore, this work provides guidance on the choice of feature extractors for future ASD research.
In future work, one might experiment with the best feature extractors from this work in conjunction with a sequence-to-sequence autoencoder. This approach would explicitly account for the temporal structure of sound and might could performance on machine types with a non-stationary sound profile.
\clearpage

\bibliographystyle{apalike}
{\small
\bibliography{bibliography}}

\end{document}